\documentclass[aps,twocolumn,superscriptaddress]{revtex4-1}
\usepackage{amsmath,amssymb,graphicx}

\usepackage[utf8]{inputenc}
\usepackage[T1]{fontenc}
\usepackage{xcolor}

\IfFileExists{newtxtext.sty}
   {\usepackage{newtxtext,newtxmath}}
   {\IfFileExists{stix.sty}
      {\usepackage{stix}}
      {\IfFileExists{mathptmx.sty}
      {\usepackage{mathptmx}}{} } }

\usepackage{textcomp}
\usepackage{bm}
\usepackage{bbold}

\IfFileExists{siunitx.sty}{\usepackage{booktabs,siunitx}}{}

\newcommand{\um}{{\textmu m}}
\newcommand{\uW}{{\textmu W}}

\usepackage{color}
\definecolor{LinkColor}{rgb}{0.15,0.3,0.5}
\usepackage{hyperref}
\hypersetup{
   pdfauthor={TS Biswas, J Xu, N Miriyala, C Doolin, T Thundat, JP Davis, KSD Beach},
   pdftitle={Time-Resolved Mass Sensing of a Molecular Adsorbate Nonuniformly Distributed Along a Nanomechnical String},
   pdfsubject={Multimode nanomechanical mass sensing},
   pdfkeywords={nanostring,} {mechanical resonator,} {silicon nitride,} {thermomechanical actuation,} {mass sensing,} {mass distribution,} {multimode analysis,} {optical detection,} {sublimation,} {RDX},
   colorlinks=true,
   citecolor=LinkColor,
   linkcolor=LinkColor,
   urlcolor=LinkColor
   }
\newcommand{\doilink}[2]{\href{http://dx.doi.org/#2}{#1}}

\begin{document}

\title{Time-Resolved Mass Sensing of a Molecular Adsorbate 
Nonuniformly Distributed\\ Along a Nanomechnical String}

\author{T. S. Biswas}
\affiliation{Department of Physics, University of Alberta, Edmonton, Alberta, Canada T6G 2E1}

\author{Jin Xu}
\affiliation{Department of Physics, University of Alberta, Edmonton, Alberta, Canada T6G 2E1}

\author{N. Miriyala}
\affiliation{Department of Chemical Engineering, University of Alberta, Edmonton, Alberta, Canada T6G 2E1}

\author{C. Doolin}
\affiliation{Department of Physics, University of Alberta, Edmonton, Alberta, Canada T6G 2E1}

\author{T. Thundat}
\affiliation{Department of Chemical Engineering, University of Alberta, Edmonton, Alberta, Canada T6G 2E1}

\author{J. P. Davis}
\email[]{jdavis@ualberta.ca}
\affiliation{Department of Physics, University of Alberta, Edmonton, Alberta, Canada T6G 2E1}

\author{K. S. D. Beach}
\email[]{kbeach@olemiss.edu}
\affiliation{Department of Physics and Astronomy, The University of Mississippi, University, Mississippi 38677, USA}

\date{June 3, 2015}

\begin{abstract} 
We show that the particular distribution of mass deposited on the 
surface of a nanomechanical resonator can be estimated by tracking the 
evolution of the device's resonance frequencies during the process of 
desorption. The technique, which relies on analytical models we have 
developed for the multimodal response of the system, enables mass 
sensing at much higher levels of accuracy than is typically achieved 
with a single frequency-shift measurement and no rigorous knowledge of 
the mass profile. We report on a series of demonstration experiments, in 
which the explosive molecule 1,3,5-trinitroperhydro-1,3,5-triazine (RDX) 
is vapor deposited along the length of a silicon nitride nanostring to 
create a dense, random covering of RDX crystallites on the surface. 
In some cases, the deposition is biased to produce distributions
with a slight excess or deficit of mass at the string midpoint.
The added mass is then allowed to sublimate away under vacuum conditions, 
with the device returning to its original state over about 4\,h 
(and the resonance frequencies, measured via optical interferometry,
 relaxing back to their pre-mass-deposition values). Our claim is that 
the detailed time trace of observed frequency shifts is rich in 
information---not only about the \emph{quantity} of RDX initially 
deposited but also about its \emph{spatial arrangement} along the 
nanostring. The data also reveal that sublimation in this case follows 
a nontrivial rate law, consistent with mass loss occurring at the exposed 
surface area of the RDX crystallites.
\end{abstract}


\maketitle

\section{Introduction}

Nanoscale mechanical resonators have proven to be useful tools for
chemical detection~\cite{Chen95,Thundat95,Arlett11} thanks to their
incredibly high sensitivity to added
mass~\cite{Yang06,Jensen08,Chaste12,Olcum14} and the ease with which
their vibrational frequencies can be measured to great accuracy. For a
device of tens or hundreds of picograms, molecules adsorbed onto the
surface at the scale of femtograms or smaller are detectable as shifts
in the resonance frequencies~\cite{Ekinci04,Ilic04a,Ilic04b}. A serious
limitation, however, is that the linear relationship between the amount
of mass added and the change in the observed resonances involves a
sometimes difficult-to-determine constant of proportionality. This fact,
which goes unacknowledged in much of the literature, is a major
impediment to high-accuracy nanomechanical mass sensing.

For a perfectly uniform distribution of added mass, the constant of 
proportionality is straightforward to compute: it is a simple geometric 
factor, given by the ratio of the resonator's mode-specific ``effective 
mass'' to its true inertial mass~\cite{Hauer13}. Even if the deposition 
is nonuniform, the situation is still manageable so long as the mass 
distribution is well characterized (and not too concentrated near the 
nodes of the detection mode). More typical of a sensing application, 
though, is that the distribution is of arbitrary form and more or less 
unknown. In that case, there is no reliable way to extract the total 
adsorbed mass from frequency-shift measurements~\cite{Naik09}, except at 
the level of an order-of-magnitude estimate.

To avoid this problem, efforts have been made to concentrate mass
adsorption to specific sites on a device through complex
fabrication~\cite{Sauer10}. An alternative approach has been to employ
multimode measurements~\cite{Hanay10,Lee11,Parkin11,Kim13,Perenon13,Wang14,Hanay15}, 
which can provide some degree of spatial resolution (limited by the
propagation of experimental uncertainties through the ``inversion
kernel''~\cite{Press07}). The simplest example is that both the size and
location of a \emph{point mass}---situated along a resonator with extent
primarily in one dimension---can be determined from a simultaneous
measurement of two resonance frequency
shifts~\cite{Dohn07,Schmid10,Stachiv12,Zhang14}. In a slightly different
context, multimode measurements have been shown to provide
single-resonator discrimination for the mass added along an array of
strongly coupled devices~\cite{Biswas14a}.

The work reported in this paper also takes advantage of a multimode 
framework (described in Sec.~\ref{SEC:Framework_for_Analysis}), and we 
find that silicon nitride nanostrings~\cite{Verbridge06,Verbridge08} are 
an excellent platform for our technique. Indeed, although the original 
experiment by Dohn and collaborators to determine the mass and location 
of a pointlike object was performed on cantilevers~\cite{Dohn07}, it 
was soon recognized that the sinusoidal mode shapes characteristic of 
nanomechanical strings greatly simplify the analysis~\cite{Schmid10}. As 
an added benefit, for materials such as stoichiometric silicon nitride, 
the high internal tension that draws a doubly clamped beam taut enough 
to become stringlike leads to correspondingly high mechanical quality 
factors. These high values aid in 
detection~\cite{Feng07,Schmid11,Suhel12} and are maintained even in the 
presence of a metallic overlayer added to the device for the purpose of 
functionalization~\cite{Biswas12,Seitner14}.
\begin{figure}[b]
\centerline{\includegraphics[width=2.9in]{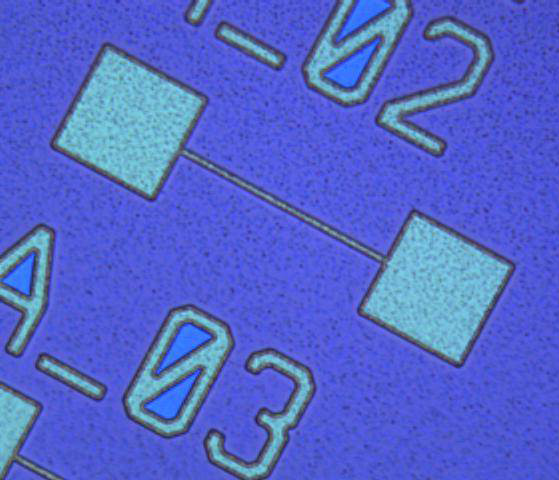}}
\caption{\label{FIG:device_image}Optical microscope image showing 
a 101.7\,\um\ nanostring with a large number of RDX crystallites 
physisorbed onto the device
surface.}
\end{figure}

In a previous study on nanostrings, real-time mass sensing was mimicked 
by carrying out sequential frequency-shift measurements of the first two 
harmonics in conjunction with pick-and-place deposition of a single 
microparticle~\cite{Schmid10}. In our work, genuine real-time 
observations are made: we measure the first, third, and fifth harmonics 
simultaneously as a function of time while mass sublimates from the 
device surface. This is possible because we are able to resolve all 
three of these modes using optical interferometry at the midpoint of the 
string with the device entirely under thermomechanical actuation. 

We show that measurements of any pair of modes reveal not only the 
instantaneous total mass of molecules adsorbed but also their 
distribution---at least in the regime where the distribution is smooth, 
slowly varying, and roughly symmetric about the resonator midpoint.
We provide explicit formulas (derived in Sec.~\ref{SEC:Framework_for_Analysis} 
and presented in Table~\ref{TAB:mass_estimates}) for the mass coefficients
of the uniform and nonuniform components of the adsorbate distribution 
in terms of all combinations of the frequency shifts measured in modes 1, 3, 
and 5---including the pseudoinverse result in which three measurements 
determine the two coefficients in the least squares sense---and we show
that the variance among these estimates can be used as an effective 
tool to judge the combined uncertainty due to measurement error and 
analytical assumptions.

The multimode measurement technique also tells us about the 
desorption characteristics of the added molecule, here the 
explosive 1,3,5-trinitroperhydro-1,3,5-triazine (RDX). In particular,
our analytical model is able to discriminate between two situations: (i) where
RDX molecules are bound to the resonator surface and thus display
first-order or Langmuir-like rate kinetics; and (ii) where the RDX 
molecules are bound to each other in crystalline formations (born from
randomly distributed nucleation sites), and sublimation occurs primarily
from the crystal interface. As we discuss in Sec.~\ref{SEC:Results},
our frequency-shift data support the latter interpretation. This is in
keeping with our expectations, since atomic force microscopy 
studies have already established that RDX forms nanoscale
crystallites on the silicon nitride surface (see the lower panel of Fig.~1
in Ref.~\onlinecite{Biswas14b}).

Our combined experimental and theoretical study of mass distribution and 
sublimation may be useful not only in the design of future explosive 
sensors~\cite{Muralidharan02,Pinnaduwage04} but also in a wide range
of nanomechanical mass-sensing applications where accurate determination of 
adsorbed molecular mass is relevant. We expect it to have particular
importance for single-molecule mass spectroscopy~\cite{Naik09}.

\section{Experimental method \label{SEC:Experimental_method}}

The nanomechanical devices under study are of a simple, doubly clamped
beam architecture: 250-nm-thick ribbons, 2--3\,{\um} wide and
100--300\,{\um} long. Their out-of-plane vibrational 
modes oscillate in the MHz range and exhibit quality factors exceeding $10^5$.
The resonators are fabricated from stoichiometric silicon nitride grown
on a sacrificial silicon dioxide layer on top of a silicon handle. They
are patterned using standard optical lithography and reactive
ion etching through the nitride and then released using a buffered
oxide etch. (A detailed description of the fabrication process is
provided in Refs.~\onlinecite{Suhel12,Biswas12,Biswas14a}.)
The resulting geometry establishes high tensile strain along
the length of the beam; as a consequence, the device acts as a classical
string, with its sinusoidal mode shapes~\cite{Suhel12} enabling the
analytical treatments that appear in
Sec.~\ref{SEC:Framework_for_Analysis}. Molecules are adsorbed onto the
nanostrings by vapor deposition of RDX that has been heated and carried
to the sample chip via a flow of nitrogen gas. It is observed that the
RDX preferentially adsorbs onto the silicon nitride and aggregates in
clusters, visible in Fig.~\ref{FIG:device_image}. We employ various
masks to gently bias adsorption around the midpoint of the string.
The weak thermal actuation of the string has no effect on
where the RDX settles.

Measurements are performed using optical interferometry. 
Light from a HeNe laser (632.8-nm wavelength) is focused onto the 
nanostrings, with part of the light reflecting from the surface of the nanostring and part
from the silicon substrate underneath the string. The interference of light therefore
encodes information on the relative separation between the nanostring and the substrate as
a modulation of the optical intensity. The laser power incident on the string is 
no more than 120\,\uW, and we have shown elsewhere~\cite{Suhel12} that there is no heating 
effect.

We focus the laser light halfway along the nanostring. The even-numbered,
out-of-plane vibrational modes have a node there, and so our measurements are 
not sensitive to them~\cite{Hauer13}. But for the odd-numbered harmonics that
we can detect, the string center is always the point of maximum amplitude; hence we are 
able to position the laser spot to high accuracy, and the signal-to-noise ratio
of the measurement is at its local maximum.

The nanostring chip is placed into an optical-access vacuum
chamber, reducing the viscous damping on the resonators~\cite{Verbridge08} and enabling
thermomechanical measurements of the nanostrings~\cite{Hauer13}. The resulting
vacuum ($\sim$\,$10^{-4}$\,torr) causes sublimation of the RDX from the nanostrings,
which is not observed under ambient conditions~\cite{Muralidharan02,Pinnaduwage04}.  
The HeNe detection laser is not resonant with any of the internal levels of RDX 
and therefore its light is not adsorbed by the molecule.

A high-frequency Zurich lock-in amplifier (model HF2LI, capable of demodulating
as many as six independent frequencies) is used to isolate up to three harmonics that
have large thermomechanical displacement at the device midpoint. We track these
harmonics as a function of time in order to extract useful information about the 
distribution of mass on the device surface and about the characteristics of the
RDX sublimation process. In practice, the continuous time trace of the optical 
interferometry measurement, $z(t)$, is used to generate a power spectral density
(PSD),
\begin{equation} \label{EQ:power_spectral_density}
S(\omega; t) = \frac{1}{T} \int_0^T\!d\tau\,e^{i\omega\tau} \int_{t-T}^{t-\tau}\!d\tau'\,z(\tau')z(\tau'+\tau),
\end{equation}
computed over a sliding time window of duration $T\approx0.85$\,s.
The ``instantaneous'' resonance frequency $\omega_n(t)$ is obtained at each time $t$ by
fitting the PSD in the vicinity of its $n$th peak to the usual (nearly-Lorentzian) damped
harmonic oscillator lineshape, 
$S(\omega; t) =  A_n \omega_n(t) / \bigl\{\bigl[\omega_n^2(t) - \omega^2\bigr]^2 + \bigl[\omega \,\omega_n(t) / Q_n\bigr]^2\bigr\}$,
with the quality factor $Q_n$ and the overall amplitude $A_n$ also optimized
as part of the fit~\cite{Suhel12,Hauer13}. This analysis is simple enough that it
can be done concurrently with the data acquisition.

\section{Framework for Analysis \label{SEC:Framework_for_Analysis}}
\subsection{Frequency shift in response to added mass}

We begin by considering the total mechanical energy (kinetic plus potential) 
of a string of length $L$ vibrating in one of its normal modes:
\begin{equation}
E_n = \int_0^L\!dx\, \mu(x) \omega_n^2 u_n^2(x).
\end{equation}
Here, $x$ measures the distance along the string; $\omega_n$ and 
$u_n(x)$ are the angular frequency and displacement profile of mode $n$;
 and $\mu(x)$ is the mass per unit length at position $x$. A small mass perturbation, arising from deposition of a distribution of molecules on
the surface of the string, leads to a modified mass distribution 
$\mu(x) \to \mu(x) + \delta\mu(x)$ and a corresponding frequency shift
$\omega_n \to \omega_n + \delta\omega_n$. To leading order, the variations 
$\delta\mu(x)$ and $\delta\omega_n$ leave the energy stationary
($dE_n = 0$); hence
\begin{equation} \label{EQ:relative_freq_shift}
-2\frac{\delta \omega_n}{\omega_n} = \frac{ \int_0^L\!dx\, \delta\mu(x) u_n^2(x) }
{ \int_0^L\!dx\, \mu(x) u_n^2(x) }.
\end{equation}
Equation~\eqref{EQ:relative_freq_shift} can be understood as a linear 
relationship $\delta \omega_n/\omega_n \propto m/M$ between the 
relative frequency shift and the ratio of the mass added, 
$m = \int_0^L\!dx\,\delta\mu(x)$, to that of the original device,
$M = \int_0^L\!dx\,\mu(x)$. The constant of proportionality
\begin{equation} \label{EQ:constant_of_proportionality}
-\frac{1}{2}\Biggl(\frac{ \int_0^L\!dx\, \delta\mu(x) u_n^2(x) }
{ \int_0^L\!dx\, \delta\mu(x)}\Biggr)
\Biggl(\frac{ \int_0^L\!dx\, \mu(x)}
{ \int_0^L\!dx\, \mu(x) u_n^2(x) }\Biggr)
\end{equation}
is unique to each mode and depends on how the resonator's own
mass and the mass adsorbed on the surface are arranged.

A variety of other effects can come into play when mass impinges on a 
resonator~\cite{Zhang13}. In devices whose vibrational modes are dominated 
by Young's modulus terms, a molecular covering can induce surface stresses
that significantly alter the bending stiffness~\cite{Dareing05} (which,
in the case of cantilevers, can itself be the basis of a detection
scheme~\cite{Berger97,Lang02}). Additional mass can also
provide a new pathway for energy dissipation, so that shifts in the resonant 
frequencies come about indirectly through changes in the mechanical quality factor.
Neither of these effects is relevant to the devices we are studying. Our experiment 
exploits two properties of nanostrings: (i) they live in a high-tension limit 
where the bending terms are negligible and the vibrational modes are almost perfectly
harmonic, and (ii) they exhibit high mechanical $Q$ that is quite robust to the
presence of any molecular overlayer.

For a high-tension string, the mode shape is a sinusoid
$u_n(x) = (2/L)^{1/2} \sin(n\pi x/L)$; and, hence
\begin{equation} \label{EQ:relative_freq_shift_string}
-2\frac{\delta \omega_n}{\omega_n} = \frac{ \int_0^L\!dx\, \delta\mu(x) 
\sin^2(n\pi x/L) }{ \int_0^L\!dx\, \mu(x) \sin^2(n\pi x/L) }.
\end{equation}
[The reflection symmetry of $u_n^2$ imposes the fundamental
limitation that $\delta \mu(x)$ cannot be distinguished from $\delta \mu(L-x)$
using frequency-shift measurements alone.]
If the unperturbed string has a uniform mass distribution $\mu(x) = M/L$,
then Eq.~\eqref{EQ:relative_freq_shift_string} specializes to 
\begin{equation} \label{EQ:relative_freq_shift_uniform}
-\frac{\delta \omega_n}{\omega_n}
= \frac{1}{M} \int_0^L\!dx\, \delta\mu(x) \sin^2\biggl(\frac{n\pi x}{L}\biggr).
\end{equation}
There are two limits worth emphasizing. In the case of a uniform mass deposition
profile, with $\delta \mu(x) = m/L$ leading to
\begin{equation}\label{EQ:naive_mass_estimate}
-\frac{\delta \omega_n}{\omega_n}
= \frac{m}{2M},
\end{equation}
the mass added $m$ can be determined
from a {\it single} frequency-shift measurement in any mode.
A strongly peaked profile represents the extreme opposite case. A point mass 
$m$ deposited at position $x_m$
leads to
\begin{equation}
-\frac{\delta \omega_n}{\omega_n}
= \frac{m}{M}  \sin^2\biggl(\frac{n\pi x_m}{L}\biggr),
\end{equation}
and hence 
the unknown values of $m$ and $x_m$
must be determined 
from {\it two} frequency-shift measurements in any pair of modes: specifically, 
\begin{align}
-\frac{m}{M} &= \frac{\delta \omega_1}{\omega_1}\biggl(1 - \frac{1}{4}\frac{\delta \omega_2/\omega_2}{\delta \omega_1/\omega_1}\biggr)^{-1},\\
x_m &= \frac{L}{\pi}\arcsin \biggl(1-\frac{1}{4}\frac{\delta \omega_2/\omega_2}{\delta \omega_1/\omega_1}\biggr)^{1/2}
\end{align}
for modes 1 and 2 (first derived in Ref.~\onlinecite{Schmid10}); and
\begin{align}
-\frac{m}{M} &= \frac{\delta \omega_1}{\omega_1}\biggl(\frac{3}{4} \pm \frac{1}{4}\frac{\delta \omega_3/\omega_3}{\delta \omega_1/\omega_1}\biggr)^{-1},\\
x_m &= \frac{L}{\pi}\arcsin \Biggl(\frac{3}{4}\pm\frac{1}{4}\sqrt{\frac{\delta \omega_3/\omega_3}{\delta \omega_1/\omega_1}}\Biggr)^{1/2}
\end{align}
for modes 1 and 3.

The general mass-sensing problem, however, is much more difficult than either 
of these two limits and has the characteristics of an ill-posed inverse problem.
The constant of proportionality in the relationship $\delta \omega_n / \omega_n \propto m/M$ is undetermined,
and there is no way to recover an arbitrary mass profile $\delta \mu(x)$, except from
an infinite number of error-free frequency-shift measurements (or, in practice, from knowledge of 
$\delta \omega_n / \omega_n$ to sufficient accuracy for all $n=1,2, 3, \ldots$ up to a cutoff corresponding
to the desired spatial resolution).

We propose to extract the information from the time evolution 
of the resonance frequencies in a handful of modes by making
some reasonable assumptions about $\delta \mu(x)$.

\subsection{Two-parameter mass distribution ansatz}

We suggest a simple, two-parameter form to represent the distribution of mass
deposited on the string:
\begin{equation} \label{EQ:two-parameter-ansatz-sinusoidal}
\delta \mu(x) = \frac{m_0}{L} + \frac{\pi m_1}{2L} \sin\bigg(\frac{\pi x}{L}\biggr).
\end{equation}
Equation~\eqref{EQ:two-parameter-ansatz-sinusoidal} is constructed so
that $m_0$ and $m_1$ have units of mass, and the total mass sitting 
on the string is 
\begin{equation}
\int_0^L\!dx\, \delta\mu(x) 
= \frac{m_0}{L}L + \frac{\pi m_1}{2L} \frac{2L}{\pi} = m_0 + m_1 \equiv m.
\end{equation}
 $m_0$ is nonnegative, but $m_1$ may be of either sign. 
$m_1 > 0$ describes a convex mass distribution, whereas $-2m_0/\pi < m_1 < 0$ describes a concave one.
The lower bound on $m_1$, which follows from $\delta\mu(L/2) \sim m_0 + \pi m_1/2  > 0$,
ensures that the mass distribution remains everywhere nonnegative.

The assumptions behind Eq.~\eqref{EQ:two-parameter-ansatz-sinusoidal}
are that (i) the distribution is symmetric under reflection about the midpoint
of the string and (ii) the distribution is smooth and slowly varying enough that it can be approximated
by one component that is uniform across the string and another that places mass preferentially 
toward (or away from, when $m_1 < 0$) its center.

\setlength{\tabcolsep}{0.3em}
\begin{table*}
 \begin{tabular}{
S[table-format = 3.2]
   S[table-format = 1.2]
      S[table-format = 3.2]
         c
            S[table-format = 1.4]
               S[table-format = 1.4]
                  S[table-format = 3.3]
                     S[table-format = 4.3]
                        S[table-format = 1.5]
                           S[table-format = 2.5]
                              S[table-format = 2.4]
                                 S[table-format = 2.4]
                                    S[table-format = 1.3]} \hline
{Length} 
   & {Width} 
      & {$M$} 
         & {Trial} 
            & {$f_1$} 
               & {$f_3$} 
                  & {$\delta f_1$} 
                     & {$\delta f_3$} 
                        & {$m_0/M$} 
                           & {$m_1/M$} 
                              & {$m\!=\!m_0\!+\!m_1$} 
                                 & $m_\text{u}$ 
                                    & $m_\text{u}/m$ \\
{(\um)} 
   & {(\um)} 
      &{(pg)} 
         &
            & {(MHz)}
               & {(MHz)}
                  & {(kHz)}
                     & {(kHz)}
                        &
                           &
                              & {(pg)}
                                 & {(pg)}
                                    & \\[0.05cm] \hline\hline 
101.07 $^\dagger$ 
  & 2.83 
     & 218.10 
        & 1
           & 2.5823 
              & 7.7928 
                 & -71.083 
                    & -210.511 
                       & 0.0506
                          & 0.00340
                             & 11.8
                                & 12.0
                                   & 1.02   \\
172.60 
   & 2.05 
      & 269.80 
         & 1 
            & 1.4871 
               & 4.4745 
                  & -0.100 
                     & -0.237 
                        & {$9.56\!\times\!10^{-6}$}
                           & {$93.7\!\times\!10^{-6}$}
                              & 0.0279
                                 & 0.0363
                                    & 1.30   \\

   &  
      & 
         & 2 
            & 1.4885 
               & 4.4785 
                  & -1.708 
                     & -11.103 
                        & 0.0139
                           & -0.00874
                              & 1.41
                                 & 0.619
                                    & 0.441  \\

   &  
      & 
         & 3 
            & 1.4898 
               & 4.4821 
                  & -13.045 
                     & -48.873 
                        & 0.0363
                           & -0.0141
                              & 5.99
                                 & 4.73 
                                    & 0.788 \\

   &
      &
         & 4
            & 1.4934
               & 4.4931
                  & -2.115
                     & -7.752
                        & 0.00554
                           & -0.00203
                              & 0.947
                                 & 0.764
                                    & 0.807 \\
216.34
   & 2.56
      & 422.30
         & 1
            & 1.2104
               & 3.6304
                  & -9.891
                     & -26.318
                        & 0.00827
                           & 0.00605
                              & 6.05
                                 & 6.90
                                    & 1.14 \\ 
309.49
   & 2.72
      & 641.88
         & 1
            & 0.8345
               & 2.5067
                  & -2.163
                     & -5.641
                        & 0.00219
                           & 0.00224
                              & 2.85
                                 & 3.33
                                    & 1.17  \\ \hline
\end{tabular}
\caption{\label{TAB:mass_values} Results of a sequence of RDX depositions onto
nanostrings spanning four different geometries, organized by string 
length and deposition trial. $M$ is the total predeposition mass
inferred from the material density and device volume. Experimentally
determined frequency values are reported in columns 5--8, viz., the
initial mode frequencies of the first and third harmonics, {$f_1$} and
{$f_3$}, and the shifts in those frequencies after the mass deposition,
{$\delta f_1$} and {$\delta f_3$}. The ratios $m_0/M$ and $m_1/M$ are
obtained from Eq.~\eqref{EQ:mass_added_modes_1_and_3}. Traditionally,
one would use the shift of the first harmonic alone in conjunction with
Eq.~\eqref{EQ:naive_mass_estimate} to determine the adsorbed mass; we
label this quantity $m_\text{u}$, since it is based on the assumption of
a uniform mass distribution. Comparison of $m_\text{u}$ with the
corresponding value $m=m_0+m_1$, coming from our improved two-component
analysis, reveals that the two estimates can differ quite substantially.
Fluctuations in the ratio $m_\text{u}/m$ indicate that the traditional
approach routinely over- or underestimates the adsorbed mass in the
range of 10\% to 50\%. This quantitative comparison demonstrates the
absolute importance of a multimode analysis for accurate mass sensing.
The dagger ($\dagger$) marks the one experiment in which the RDX
deposition process is completely unbiased; the others involve some
degree of masking in order to engineer a nonuniform mass distribution.}
\end{table*}

Putting Eq.~\eqref{EQ:two-parameter-ansatz-sinusoidal} into
Eq.~\eqref{EQ:relative_freq_shift_string} gives an expression for the
frequency shifts in the various modes: 
\begin{equation} \label{EQ:freq_shifts_all_modes}
-\frac{\delta \omega_n}{\omega_n} = \frac{1}{M} \biggl( \frac{m_0}{2} + \frac{2n^2m_1}{4n^2-1}\biggr).
\end{equation}
Inverting the relationship, we find that the values of $m_0$ and $m_1$ can be estimated
from frequency-shift measurements on any pair of modes. For instance, in the case of 
modes 1 and 2 and modes 1 and 3, we find that
\begin{alignat}{2} \label{EQ:mass_added_modes_1_and_2}
\frac{m_0}{M} &= 8\frac{\delta \omega_1}{\omega_1} - 10 \frac{\delta \omega_2}{\omega_2},
&\frac{m_1}{M}&= \frac{15}{2}\biggl( \frac{\delta \omega_2}{\omega_2} - 
\frac{\delta \omega_1}{\omega_1}\biggr); \\
\label{EQ:mass_added_modes_1_and_3}
\frac{m_0}{M} &= \frac{27}{4}\frac{\delta \omega_1}{\omega_1} 
- \frac{35}{4}\frac{\delta \omega_3}{\omega_3},& \quad
\frac{m_1}{M} &= \frac{105}{16}\biggl( \frac{\delta \omega_3}{\omega_3} - 
\frac{\delta \omega_1}{\omega_1}\biggr).
\end{alignat}
According to Eq.~\eqref{EQ:freq_shifts_all_modes}, higher modes (larger $n$)
exhibit decreasing differentiation in the $m_1$ coefficient and so become
less useful for distinguishing the components of the mass distribution,
$-\delta \omega_n/\omega_n \to (m_0 + m_1)/2M$, independent of $n$,
as $n\to\infty$.

A revealing, preliminary application of this style of analysis is provided in
Table~\ref{TAB:mass_values}, where Eq.~\eqref{EQ:mass_added_modes_1_and_3}
is enlisted to characterize the mass distributed on seven nanostrings of four 
different lengths following various levels of RDX exposure.
In each instance, the resonant frequencies of modes 1 and 3 are measured
before and after RDX vapor deposition. We find that the estimates for $m_1$
exhibit both positive and negative sign (meaning convex and concave mass profiles), 
and in some cases show magnitude $|m_1|$ comparable to $m_0$ itself.
This is evidence for a quite substantial variation in how RDX settles along the
length of the string from one experiment to the next, and it confirms that a mass
distribution close to uniform is only rarely the outcome of our sample preparation.
The results in Table~\ref{TAB:mass_values} are fully consistent with our intentional 
biasing of the vapor-deposition process.

How confident should we be in this analysis? 
One issue that arises is how to quantify the uncertainty arising from our
choice of ansatz.
The coefficients appearing in Eqs.~\eqref{EQ:mass_added_modes_1_and_2}
and \eqref{EQ:mass_added_modes_1_and_3} are specific to our choice 
of the sine function in Eq.~\eqref{EQ:two-parameter-ansatz-sinusoidal}. As a test
of the robustness of our assumption, it will be helpful to check the results against an
alternative functional form. We therefore also try a symmetric polynomial 
\begin{equation} \label{EQ:two-parameter-ansatz-polynomial}
\delta\mu(x) = m_0 + 6m_1\frac{x}{L}\biggl(1-\frac{x}{L}\biggr).
\end{equation}
It is also important is to look for consistency between the results
achieved with measurements on different sets of modes.
As discussed in Sec.~\ref{SEC:Experimental_method},
the laser in our optical detection system is parked at the string midpoint, so 
only the odd-numbered modes are measured. Table~\ref{TAB:mass_estimates}
lists all the possible mass determinations using frequency shifts in modes 1, 3, 
and 5 (exact determinations with the three available pairs and one
overdetermination with the full triplet of modes), 
under the assumption of a mass profile following either
Eq.~\eqref{EQ:two-parameter-ansatz-sinusoidal} or
Eq.~\eqref{EQ:two-parameter-ansatz-polynomial}.

A reasonable final value of the fractional mass added comes 
from averaging those eight estimates:
\begin{equation} \label{EQ:mass_mean_estimate}
\begin{split}
\frac{m}{M} &= \frac{m_0 + m_1}{M}\\
&= 0.108\,031\frac{\delta\omega_1}{\omega_1} 
- 0.534\,596 \frac{\delta\omega_3}{\omega_3}  - 1.573\,434 \frac{\delta\omega_5}{\omega_5}.
\end{split}
\end{equation}
The total uncertainty, $\bigl(\Delta_{e}\bigl[m/M\bigr]^2 + \Delta_{a}\bigl[m/M\bigr]^2\bigr)^{1/2}$,
is taken to be a combination of the experimental errors arising from imperfect knowledge of the relative frequency shifts,
\begin{multline} \label{EQ:error_shift}
\Delta_{e}\biggl[\frac{m}{M}\biggr]^2 = 0.0117 \Delta_{e}\biggl[\frac{\delta\omega_1}{\omega_1}\biggr]^2 
+ 0.2858\Delta_{e}\biggl[\frac{\delta\omega_3}{\omega_3}\biggr]^2 \\
+ 2.4757 \Delta_{e}\biggl[\frac{\delta\omega_5}{\omega_5}\biggr]^2, 
\end{multline}
and the uncertainty due to the choice of ansatz, which we approximate by the spread around
the mean estimate given in Eq.~\eqref{EQ:mass_mean_estimate}:
\begin{multline} \label{EQ:error_ansatz}
\Delta_{a}\biggl[\frac{m}{M}\biggr]^2  = 0.007\,91\biggl(\frac{\delta\omega_1}{\omega_1}\biggr)^2
+ 1.7344\biggl(\frac{\delta\omega_3}{\omega_3}\biggr)^2\\
+ 1.5138\biggl(\frac{\delta\omega_5}{\omega_5}\biggr)^2 
 -0.2285\frac{\delta\omega_1}{\omega_1}\frac{\delta\omega_3}{\omega_3}\\
+0.2127\frac{\delta\omega_1}{\omega_1}\frac{\delta\omega_5}{\omega_5}
-3.2403\frac{\delta\omega_3}{\omega_3}\frac{\delta\omega_5}{\omega_5}.
\end{multline}
The notation $\Delta_e[\delta\omega_n / \omega_n]$ in Eq.~\eqref{EQ:error_shift} is meant 
to indicate the error in the relative frequency shift, understood as the ratio of two inexact quantities
computed according to 
$\Delta_e[\delta\omega_n / \omega_n]^2 = (\Delta_e[\delta\omega_n]/\omega_n)^2 + (\Delta_e[\omega_n]\delta\omega_n/\omega_n^2)^2$.

\setlength{\tabcolsep}{1.2em}
\begin{table*}
\begin{center}
\begin{tabular}{cccc} \hline
Ansatz & Modes & $m_0/M$ & $m_1/M$\\[0.1cm] \hline\hline && \\[-0.25cm]
Eq.~\eqref{EQ:two-parameter-ansatz-sinusoidal} &1, 3 & 
$\frac{27}{4}\frac{\delta\omega_1}{\omega_1} - \frac{35}{4}\frac{\delta\omega_3}{\omega_3} $ &
$\frac{105}{16}\bigl(-\frac{\delta\omega_1}{\omega_1} + \frac{\delta\omega_3}{\omega_3}\bigr) $
\\[0.2cm]
&1, 5 &
$\frac{25}{4}\frac{\delta\omega_1}{\omega_1} - \frac{33}{4}\frac{\delta\omega_5}{\omega_5} $ &
$\frac{99}{16}\bigl(-\frac{\delta\omega_1}{\omega_1} + \frac{\delta\omega_5}{\omega_5}\bigr) $
\\[0.2cm]
&3, 5 &
$\frac{875}{8}\frac{\delta\omega_3}{\omega_3} - \frac{891}{8}\frac{\delta\omega_5}{\omega_5} $ &
$\frac{3465}{32}\bigl(-\frac{\delta\omega_3}{\omega_3} + \frac{\delta\omega_5}{\omega_5}\bigr) $
\\[0.2cm]
&1, 3, 5 &
$\frac{15\,007}{2318}\frac{\delta\omega_1}{\omega_1} - \frac{36\,365}{37\,088}\frac{\delta\omega_3}{\omega_3}
- \frac{128\,205}{37\,088}\frac{\delta\omega_5}{\omega_5} $ &
$-\frac{58\,905}{9272}\frac{\delta\omega_1}{\omega_1} + \frac{107\,415}{9272}\frac{\delta\omega_3}{\omega_3}
+ \frac{42\,207}{9272}\frac{\delta\omega_5}{\omega_5} $\\[0.2cm] \hline && \\[-0.25cm]
Eq.~\eqref{EQ:two-parameter-ansatz-polynomial}
 &1, 3 & 
$\frac{1+3\pi^2}{4}\frac{\delta\omega_1}{\omega_1} - \frac{9+3\pi^2}{4}\frac{\delta\omega_3}{\omega_3} $ &
$\frac{3\pi^2}{4}\bigl(-\frac{\delta\omega_1}{\omega_1} + \frac{\delta\omega_3}{\omega_3}\bigr) $
\\[0.2cm]
&1, 5 &
$\frac{3+25\pi^2}{36}\frac{\delta\omega_1}{\omega_1} - \frac{25(3+\pi^2)}{36}\frac{\delta\omega_5}{\omega_5} $ &
$\frac{25\pi^2}{36}\bigl(-\frac{\delta\omega_1}{\omega_1} + \frac{\delta\omega_5}{\omega_5}\bigr) $
\\[0.2cm]
&3, 5 &
$\frac{9+75\pi^2}{8}\frac{\delta\omega_3}{\omega_3} - \frac{25(1+3\pi^2)}{8}\frac{\delta\omega_5}{\omega_5} $ &
$-\frac{75\pi^2}{8}\bigl(-\frac{\delta\omega_3}{\omega_3} + \frac{\delta\omega_5}{\omega_5}\bigr)$
\\[0.2cm]
&1, 3, 5 &
$\frac{217 + 975\pi^2}{1358}\frac{\delta\omega_1}{\omega_1} - \frac{5607+1725\pi^2}{5432}\frac{\delta\omega_3}{\omega_3}
- \frac{6125+2175\pi^2}{5432}\frac{\delta\omega_5}{\omega_5} $ &
$\frac{75\pi^2}{5432}\bigl(-\frac{52\delta\omega_1}{\omega_1} + \frac{23\delta\omega_3}{\omega_3}
+ \frac{29\delta\omega_5}{\omega_5}\bigr) $ \\[0.2cm] \hline
\end{tabular}
\end{center}
\caption{\label{TAB:mass_estimates}Estimates of the uniform ($m_0$) and nonuniform ($m_1$) mass components
as a function of the relative frequency shifts. A solution to the system of equations is 
given for each pair of modes; also shown is the least-squares, pseudoinverse solution 
involving all three modes.}
\end{table*}

\subsection{Sublimation model}

The preliminary results appearing in Table~\ref{TAB:mass_values} are obtained from
two discrete measurements of the devices' resonance frequencies, before and after RDX 
exposure. We have the capability, however, to make ongoing measurements of the 
resonance frequencies; these change continuously in time as the adsorbate molecules are removed
from the nanostring because of the vacuum environment. The main thrust of this paper is
to show how we can exploit this much richer data set.

Therefore it is useful to sketch out a model of how the mass on the string evolves with time
in our experimental setup. The basic assumption is that the molecules are either residing on the 
string ($m$) or existing as vapor in the chamber ($m_\text{v}$). RDX sublimates from the string 
surface at a uniform rate $\alpha$, and there is an incoming flux of molecules returning to 
the surface due to intermolecular collisions, denoted by $\beta$. Finally, there is $\gamma$, the
rate at which the chamber is being evacuated. This picture leads to a coupled pair of 
rate equations:
\begin{equation}
\frac{d}{dt}\begin{pmatrix} m \\ m_\text{v} \end{pmatrix}
= \begin{pmatrix} -\alpha & \beta \\
\alpha & -(\beta + \gamma) \end{pmatrix}
\begin{pmatrix} m \\ m_\text{v} \end{pmatrix}.
\end{equation}
As a consequence, there are two rate constants, given by the
eigenvalues of the matrix. When $\gamma = 0$, the
eigenvalues are $\lambda_1 = 0$ and $\lambda_2 = \alpha + \beta$;
hence, the system equilibrates at a rate $\alpha + \beta$ to 
a steady state with $m_\text{v}/m = \alpha/\beta$. On the other hand,
when $\gamma$ is a fast rate (i.e., larger than both $\alpha$ and
$\beta$, which is the experimentally relevant situation) we 
find that
\begin{equation}
\begin{split}
\lambda_1 &= -\alpha + \frac{\alpha \beta}{\gamma} +  \frac{\alpha \beta(\alpha - \beta)}{\gamma^2} + O(\gamma^{-3}),\\
\lambda_2 &= -\gamma - \beta - \frac{\alpha \beta}{\gamma} -  \frac{\alpha \beta(\alpha - \beta)}{\gamma^2} + O(\gamma^{-3}).
\end{split}\end{equation}
The full time dependence of $m$ and $m_\text{v}$ is then given by
\begin{multline}
\begin{pmatrix} m(t) \\ m_\text{v}(t) \end{pmatrix} = \frac{\gamma m(0) + \beta m_\text{v}(0)}{\gamma^2+\alpha\beta}
\begin{pmatrix} \gamma \\ \alpha \end{pmatrix}e^{-(\alpha - \alpha \beta/\gamma) t}\\
+ \frac{\gamma m_\text{v}(0) - \alpha m(0)}{\gamma^2+\alpha\beta}
\begin{pmatrix} -\beta \\ \gamma \end{pmatrix}e^{-(\gamma + \beta + \alpha \beta/\gamma)t}.
\end{multline}
For times $t \gg (\gamma + \beta + \alpha\beta/\gamma)^{-1}$, the mass on the
string decays according to
\begin{equation}
m(t) = \frac{m(0) + (\beta/\gamma)m_\text{v}(0)}{1+\alpha\beta/\gamma^2} 
e^{-\alpha(1-\beta/\gamma)t}.
\end{equation}
Moreover, if the chamber is being very aggressively evacuated, then the behavior
looks like 
\begin{equation} \label{EQ:simple_decay}
m(t) = m(0)e^{-\alpha_\text{eff} t},
\end{equation}
with $\alpha_\text{eff} = \alpha - \alpha\beta/\gamma \approx \alpha$ close
to the intrinsic sublimation rate of the adsorbate, and we can safely proceed as
if $m_\text{v} \approx 0$ at all times after the pump is activated.

A refinement of this model is to consider the possibility that the rate of mass loss
goes as a fractional power of the current mass load. This may be appropriate here
since the RDX on the surface is known to aggregate, rather than arranging in a
smooth monolayer. Sublimation in that case is likely to be
at least partially controlled by loss from the surface area of the RDX crystallites
($A \sim m^{2/3}$), which puts a lower bound of 2/3 on the effective scaling exponent. 
If the distribution of crystallites along the device is roughly uniform, then we can account
for this situation as follows: 
\begin{equation}
\frac{dm}{dt} = -\alpha m_\star (m/m_\star)^\phi.
\end{equation}
This description requires that we introduce a new material-specific mass scale $m_\star$
and a phenomenological exponent $\phi$.
We anticipate a value $2/3 \le \phi \le 1$, with conventional exponential decay
recovered at the upper end of that range. When $\phi \neq 1$, we find that
\[ \frac{dm}{m^\phi} = \frac{1}{1-\phi}d(m^{1-\phi})= -\alpha m_\star^{1-\phi} dt. \]
This leads to
\begin{equation}\label{EQ:powerlaw_decay}
m(t) = m(0)\biggl[ 1 - \frac{t}{t_\text{r}} \biggr]^{1/(1-\phi)},
\end{equation}
where, $t_\text{r} = [m(0)/m_\star]^{1-\phi}/\alpha(1-\phi)$
is the finite removal time after which all RDX has left the nanostring.

We now emphasize an important feature that distinguishes 
between the two kinds of behavior. When $\phi = 1$, the
decay is exponential at a constant rate
\begin{equation} \label{EQ:const_decay_rate}
-\frac{1}{t}\ln\frac{m(t)}{m(0)} = \alpha. 
\end{equation}
For a nontrivial value of the exponent, however, 
the mass loss is characterized by an apparent decay rate that increases over time:
\begin{equation}\label{EQ:drifting_decay_rate}
\begin{split}
 -\frac{1}{t}\ln\frac{m(t)}{m(0)} 
 &= -\frac{1}{(1-\phi)t}\ln \biggl[ 1 - \frac{t}{t_\text{r}} \biggr] \\
 &= \frac{1}{(1-\phi)t_\text{r}}\biggl[1 + \frac{t}{2t_\text{r}}
 + \frac{t^2}{3t_\text{r}^2} + \cdots \biggr]\\
 &= \alpha \biggl(\frac{m_\star}{m(0)}\biggr)^{1-\phi}\biggl[1 + 
 \frac{(1-\phi)\alpha t}{2} \biggl(\frac{m_\star}{m(0)}\biggr)^{1-\phi} \\
 &\qquad+ \cdots \biggr].
 \end{split}\end{equation}

Finally, a more complete description must account for a mass distribution
that varies along the length of the string. We treat the deposited 
mass $\delta\mu(x,t)$ as a space- and time-dependent field subject
to the governing equation
\begin{equation}
\frac{\partial}{\partial t} \delta\mu(x,t) = - \alpha\mu_\star 
\biggl(\frac{\delta\mu(x,t)}{\mu_\star}\biggr)^\phi.
\end{equation}
Here, $\mu_\star$ is a stand-in for $m_\star/L$.
In the previously considered situations, where either $\phi = 1$ 
or the initial mass distribution is uniform, the mass distribution 
simply shrinks away while preserving its overall shape.
That is not generally true:
\begin{equation} \label{EQ:mass_profile}
\delta\mu(x,t) = \Bigl[\delta\mu(x,0)^{1-\phi} -(1-\phi)\alpha  \mu_\star^{1-\phi}t
\Bigr]^{1/(1-\phi)}.
\end{equation}
Equation~\eqref{EQ:mass_profile} also makes clear that, unlike in Eq.~\eqref{EQ:powerlaw_decay},
$\delta \mu$ reaches zero at a removal time that is different at each point along the string.

With Eq.~\eqref{EQ:mass_profile} in hand, it is straightforward to obtain
the time-dependent total mass
\begin{equation} \label{EQ:total_mass_integrated}
m(t) = \int_0^L \!dx\,\delta \mu(x,t)
\end{equation}
or any of the relative frequency 
shifts
\begin{equation} \label{EQ:relative_freq_shifts_time_dependence}
-\frac{\delta\omega_n(t)}{\omega_n} = \frac{1}{M} \int_0^L \!dx\, \delta\mu(x,t) 
\sin^2 \biggl(\frac{n \pi x}{L} \biggr).
\end{equation}
The latter is simply Eq.~\eqref{EQ:relative_freq_shift_uniform} with $\delta \mu(x)$ 
replaced by $\delta \mu(x,t)$.

\begin{figure}[b]
\begin{center}
\includegraphics[width=3.1in]{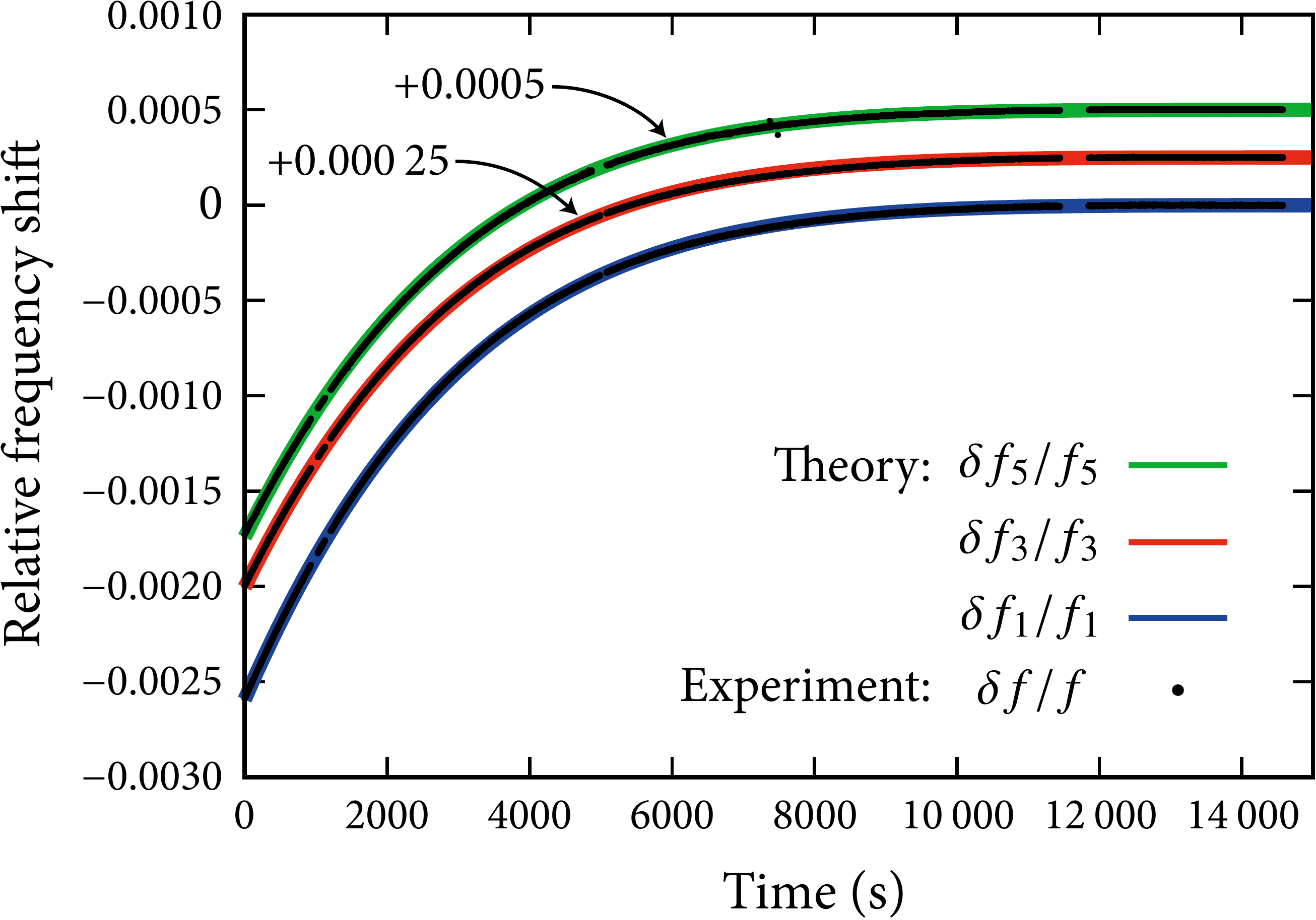}
\includegraphics[width=3.1in]{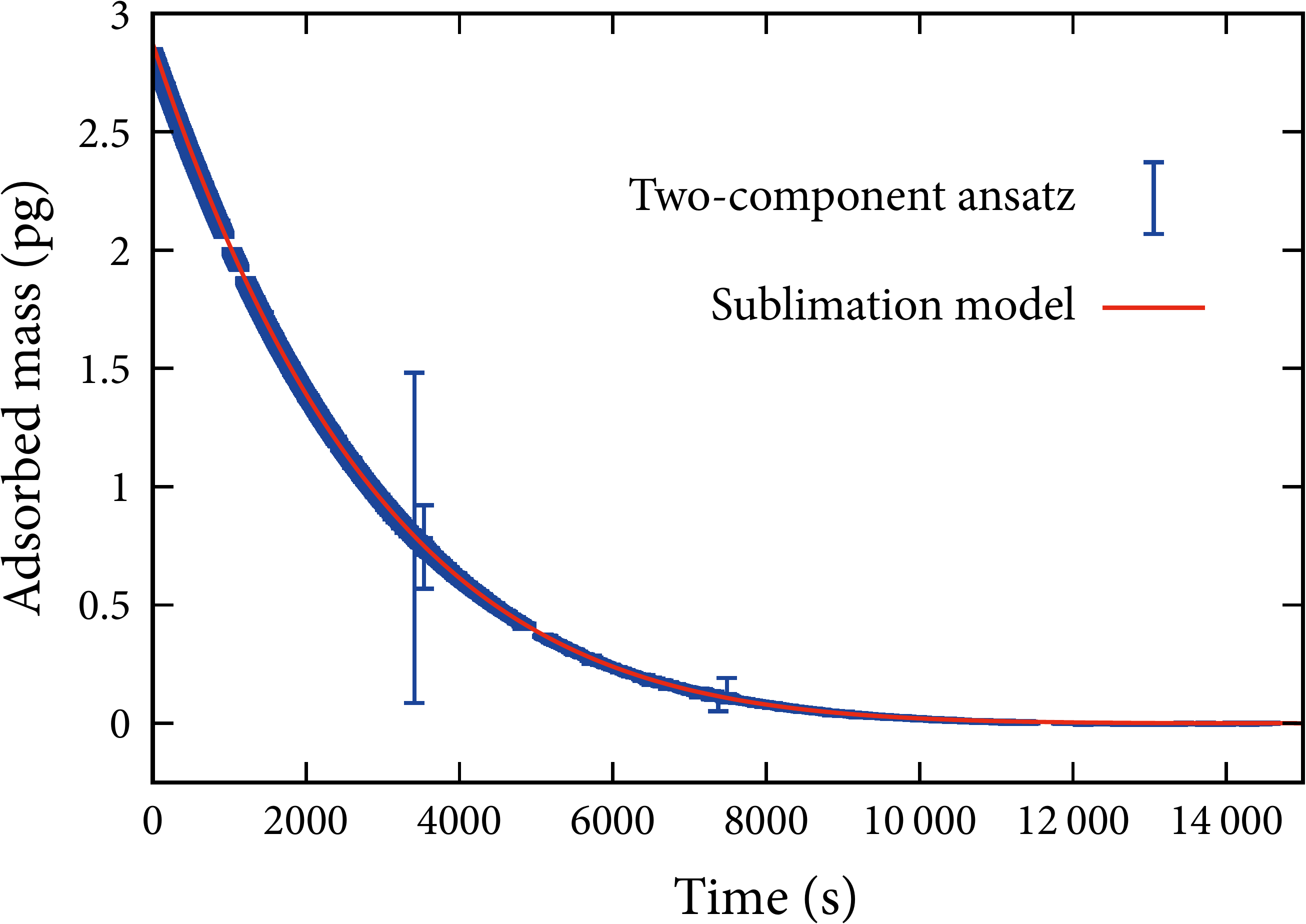}
\end{center}
\caption{\label{FIG:shifts_and_mass}Multimode frequency shifts as a function of
elapsed time and the resulting RDX mass extraction. In the upper panel, each 
experimental data point represents a relative shift in the resonance frequency
of one of modes 1, 3, or 5, obtained from the finite-time-window PSD [as per 
Eq.~\eqref{EQ:power_spectral_density}]. (The time-window resolution is fine 
enough that the experimental points bleed together into nearly continuous curves.)
For visual clarity, the mode 3 and mode 5 data sets are translated upward by 0.000\,25 
and 0.0005, respectively. The three solid lines are the result of
matching our theoretical model  [viz.,\ Eq.~\eqref{EQ:relative_freq_shifts_time_dependence}
in combination with Eqs.~\eqref{EQ:mass_profile} and \eqref{EQ:three_parameter_ansatz}]
to the experimental observations. The lower panel shows the corresponding mass.
Two estimates are presented: one is a discrete set of values calculated at each
instant according to Eqs.~(\ref{EQ:mass_mean_estimate}--\ref{EQ:error_ansatz});
the other is a continuous curve [produced numerically with 
Eq.~\eqref{EQ:total_mass_integrated}] based on the global fit achieved in the
upper panel. The two estimates are found to be in excellent agreement.}
\end{figure}

\section{Results \label{SEC:Results}}

We now apply the models developed in Sec.~\ref{SEC:Framework_for_Analysis}
to our experimentally acquired frequency shifts. The results reported here are taken from
measurements on a single device of length 310\,\um\ and mass 641.88\,pg. The device's first,
third, and fifth modes are tracked over the course of 4\,h.

Our approach is to match Eq.~\eqref{EQ:relative_freq_shifts_time_dependence}, for the cases of 
$n=1,3,5$, to our time traces of the resonance frequencies. The integrals are carried out numerically
with Eq.~\eqref{EQ:mass_profile} serving as the model for $\delta \mu(x,t)$.
In the spirit of our original two-component ansatz, we choose to express the initial mass distribution
in this slightly more expressive form:
\begin{equation} \label{EQ:three_parameter_ansatz}
\delta \mu(x,0) = \frac{m_0}{L} + \frac{m_1}{L} N(p)
\biggl[\frac{x}{L}\biggl(1-\frac{x}{L}\biggr)\biggr]^p.
\end{equation}
The exponent $p$ adds some additional flexibility with regard to the shape
of the nonuniform contribution [and permits a rough interpolation
between Eqs.~\eqref{EQ:two-parameter-ansatz-sinusoidal}
and \eqref{EQ:two-parameter-ansatz-polynomial}]. The normalization factor
\[ N(p) = \frac{2^{1+2\theta}}{\sqrt{\pi}}
\frac{\Gamma(3/2+p)}{\Gamma(1+p)} \]
is chosen to preserve the property that $m = \int\!dx\,\delta\mu(x,0) = m_0 + m_1$.

The set of independent variational parameters $m_0$, $m_1$, $p$, $\phi$, and 
$C = (1-\phi)\alpha \mu_\star^{1-\phi}$ is simultaneously optimized using
the nonlinear least-squares Levenberg-Marquardt algorithm~\cite{Levenberg44,Marquardt63}.
The number of degrees of freedom---only five---is remarkably small relative to the
large number ($\approx 3 \times 7000$) of data points.
We find that the total mass on the string is $m_0+m_1 = 2.871(2)$\,pg
and that the decay exponent has a value $\phi=0.826(6)$.
The resulting high-quality fit is displayed in the upper panel of Fig.~\ref{FIG:shifts_and_mass}.
The lower panel of the same figure shows the mass extracted at each instant from
frequency-shift measurements according to Eq.~\eqref{EQ:mass_mean_estimate},
with error bars on the data points obtained from the quadratic mean of Eqs.~\eqref{EQ:error_shift}
and \eqref{EQ:error_ansatz}.
There are a handful of points in the plot where the uncertainty estimate
blows up, but this amounts to just a few blips in the more than 7000 data points. The
solid line superimposed on the data points is produced by integrating [via Eq.~\eqref{EQ:total_mass_integrated}]
the evolving mass profile that emerges from the fitting procedure [Eq.~\eqref{EQ:mass_profile} with globally optimized
parameters]. The consistency between these two approaches gives us confidence that our
estimate of the total adsorbed mass is highly reliable.

 \begin{figure}
\begin{center}
\includegraphics[width=3.1in]{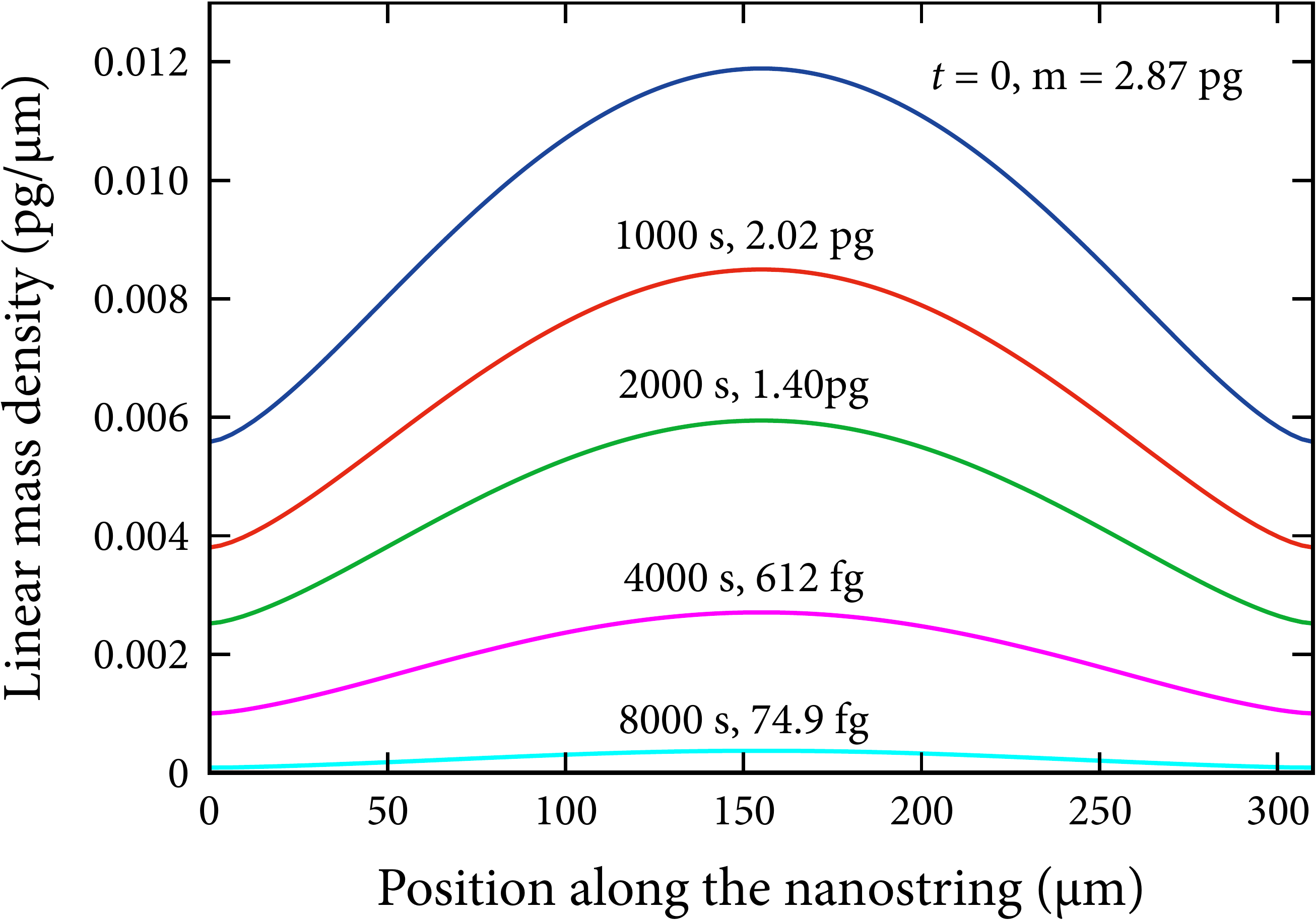}
\end{center}
\caption{\label{FIG:mass_profiles}  
Profiles of the mass distributed along the nanostring 
of length 310\,\um\ and mass 641.88\,pg at representative
moments in time. Shown here are plots of Eq.~\eqref{EQ:mass_profile} 
evaluated with optimized variational parameters at the appropriate
time $t$. Accompanying each distribution is a label giving the 
time of the snapshot and integrated weight under the curve.
}
\end{figure}

\begin{figure}
\begin{center}
\includegraphics[width=3.1in]{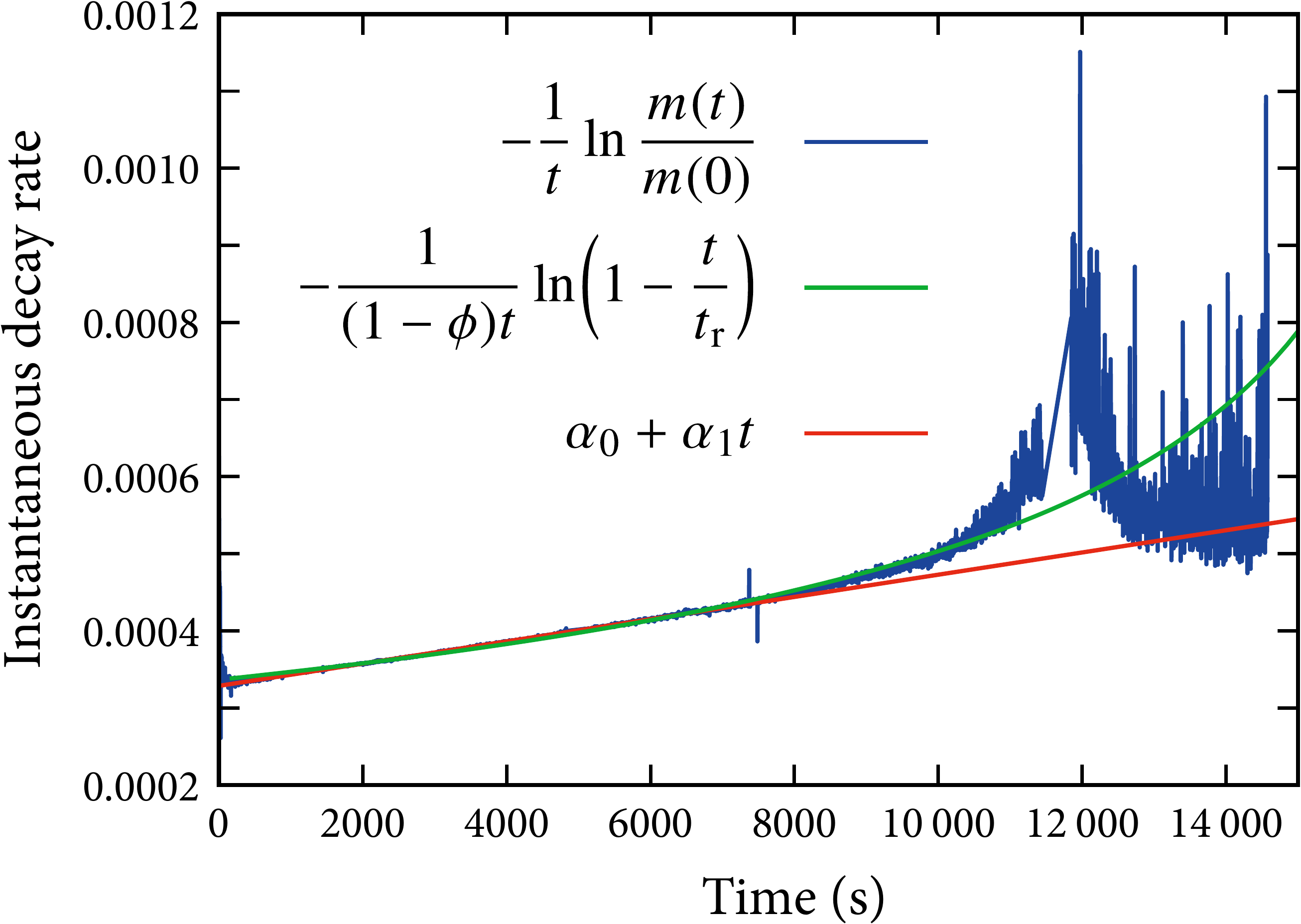}
\end{center}
\caption{\label{FIG:decay_rate}  
The instantaneous rate at which RDX desorbs from the nanostring
clearly increases over time. The data shown are a logarithmic rescaling
of the data appearing in the lower panel of Fig.~\ref{FIG:shifts_and_mass}.
A linear fit over the time interval [250\,s, 7750\,s] produces
$\alpha(t) = \alpha_0 + \alpha_1t = 0.000\,328\,65(6) + 1.445(1) \times t$.
Substantially better agreement is achieved with a function having
the form of Eq.~\eqref{EQ:drifting_decay_rate} with exponent
$\phi = 0.8278$ (fully consistent with the value $\phi=0.826(6)$ used 
in the upper panel of Fig.~\ref{FIG:shifts_and_mass}) 
and removal time $t_\text{r} = 17\,247$\,s.
}
\end{figure}

Figure~\ref{FIG:mass_profiles} shows snapshots of the mass profile at 
various times in the experiment. Such a determination of the real-time
mass distribution may have important applications, such as the study 
of diffusion of molecules along the surface~\cite{Yang11}, or atomic 
layer reconstruction~\cite{Tavernarakis14}, of a nanomechanical resonator. 

Figure~\ref{FIG:decay_rate} indicates the effective instantaneous
decay rate, calculated as per Eqs.~\eqref{EQ:const_decay_rate} and 
\eqref{EQ:drifting_decay_rate}.
One can see clearly that the rate is not
constant in time---and hence it is incompatible with pure
exponential decay. Rather, it seems to increase steadily.
Moreover, between 8000\,s and 11\,000\,s, it turns up in a way that
is consistent with the fractional power of $\phi$ found
in our fit to the frequency-shift measurements.
It is worth reiterating that this implies that sublimation 
occurs from the surface of crystallites---and not uniformly
from the surface of the device---in agreement with optical 
(see Fig.~\ref{FIG:device_image}) and atomic force 
(see Ref.~\onlinecite{Biswas14b}) microscopy.

Something further we have checked is whether the fitting can be meaningfully 
improved by allowing $\alpha \to \alpha(x)$ to vary along the length of the string. One might
imagine, for instance, that the sublimation rate has a uniform contribution controlled  by the partial
pressure of RDX in the evacuated chamber and a locally varying component 
that comes about in some unspecified way from the influence of the mechanical motion.
We have carefully considered various {\it ad hoc} models, but we find consistently that allowing spatial variation
in $\alpha(x)$ does not improve the fit enough to justify the additional variational degrees of freedom~\cite{Akaike74}.

\section{Conclusions}

Taking advantage of the sinusoidal mode shape of silicon nitride nanostrings 
under high tension, we develop analytical models for the multimode
frequency shifts that occur as a result of mass deposited onto such devices.
We apply those models to real resonators in the lab and demonstrate our
ability to make reliable estimates of the total amount and distribution of 
adsorbed molecules. This work provides experimental determination of a 
nonuniform and nonpointlike mass distribution.

Our multimode technique produces reliable, time-evolving
estimates of the total mass adsorbed on the resonator with a sensitivity 
on the order of a few femtograms, even in the absence of detailed knowledge 
of the initial mass distribution. The time trace of frequency measurements in
a handful of low-lying normal modes is sufficient to reconstruct
that missing information and thus fix the otherwise unknown constant of 
proportionality [appearing as Eq.~\eqref{EQ:constant_of_proportionality}].

Notably, the estimate of the mass added, as determined by our analysis, is 
often found to disagree with the value arrived at by the more traditional approach,
which relies on a frequency shift in a single mode and incorporates no understanding
of how the added mass is arranged on the device. This reveals the extreme importance 
of such an analysis to accurate mass sensing with nanomechanical resonators. 

In addition, because the adsorbed molecules sublimate from the surface of the 
nanostrings in our experimental system, we have had an opportunity to analyze
their desorption characteristics. Real-time measurements allow for significantly 
improved estimates of the mass distributions and reveal the evolution of those
distributions over time. This may prove a useful tool in studying molecular 
diffusion~\cite{Yang11} or the formation of monolayers onto nanomechanical 
resonators~\cite{Tavernarakis14}. 

Finally, multimode analysis also leads to insights into the form in which the 
molecules are removed from the nanostring surface by the vacuum. Specifically, 
we are able to develop sublimation models that account for uniform-rate loss
from the surface and local-mass-dependent loss from the surface area of a 
crystallite. Only the latter is consistent with the observed frequency shifts 
and the microscopy of the devices. This may have important applications in 
real-world sensing of explosive molecules such as RDX, for example in airport 
passenger and luggage screening. 

\begin{acknowledgments}

The authors acknowledge support from the Natural Sciences and Engineering Research Council of Canada; 
the Canada Foundation for Innovation; Alberta Innovates Technology Futures; Grand Challenges Canada; 
Alberta Innovates Health Solutions; and the Office of Research and Sponsored Programs of the University 
of Mississippi. One of the authors (K.S.D.B.) benefitted from a stay at the Aspen Center for Physics under 
NSF Grant No.\ 1066293.

\end{acknowledgments}

\end{document}